\documentstyle[12pt,twoside,fleqn,epsfig,espcrc1]{article}

\title{\vspace{-2cm}{\bf Triton photodisintegration with realistic potentials}\thanks{
Talk given at the XVth International Conference on Few-Body Problems in
Physics (22-26 July 1997, Groningen, The Netherlands)}}

\author{\underline{W. Schadow} and W. Sandhas\thanks{This work was supported
    by the Deutsche Forschungsgemeinschaft under Grant No. Sa 327/23-1} \\
  \vspace{3mm} Physikalisches Institut der Universit{\"a}t Bonn, Endenicher
  Allee 11-13, D-53115 Bonn, Germany}

\begin{document}
\vspace{-1cm}

\maketitle

\begin{abstract}
The photodisintegration of $^{3}$H is treated by means of
coupled integral equations using separable versions of the Paris and
the Bonn potentials in their kernel. The differential cross section
for the inverse reaction is obtained via detailed balance.
For the latter process good agreement with the data is found when
including final-state interaction, meson exchange currents, higher
partial waves in the potential, and electric quadrupole contributions
in the electromagnetic interaction.
\end{abstract}

\section{INTRODUCTION}

We present triton photodisintegration results obtained for the Paris, the Bonn
{\em A} and Bonn {\em B} potentials. Instead of the {\em W}-matrix representation
\cite{Bart86} employed in our previous calculations \cite{Schad97a}, we now
use the Ernst-Shakin-Thaler (EST) expansion \cite{Erns73} of these
two-body inputs: PEST, BAEST and BBEST \cite{Haid84,Haid86a}.  
As in Ref.
\cite{Schad97a} we are interested in the $ \gamma + t \rightarrow n + d$ cross
section over an energy range from threshold up to 40 MeV, while other recent
investigations are focused primarily on polarization observables at some
specific energies \cite{Jourd86,Ishi92a,Fons95a}. We confirm in this way the
strong potential dependence in the peak region observed in 
Ref. \cite{Schad97a}, as well as the importance of higher subsystem
partial waves at energies above the peak region. There the fully
converged results obtained in the present calculations lie between
the two competing sets of data \cite{Kosi66,Skop81}. A strong correlation
of the peak heights and the triton binding energies is found. Making
use of detailed balance we calculate also the differential cross section
for the radiative capture   process $p + d \rightarrow \gamma + \, ^{3}$He.
Good agreement between theory and experiment is achieved at
12.1 MeV and 15.3 MeV, when incorporating
final-state interaction and $p$-waves in the potential.

\section{THEORY}

The Alt-Grassberger-Sandhas (AGS)) equations are well known to go over into
effective two-body Lippmann-Schwinger equations \cite{Alt67} when representing
the input two-body {\em T}-operators in separable form. The neutron-deuteron
scattering amplitude, thus, is determined by

\begin{equation}
\label{scampl}
 {\cal T} (\vec q, \vec q\,'') = {\cal V} (\vec q, \vec q\, '')
 + \int \! d^{\,3} q' \: {\cal V} (\vec q, \vec q\,')\:{\cal G}_0 (\vec q\,')
\: {\cal T} (\vec q\,',\vec q\,'') .
\end{equation}

\noindent
Applying the same technique to the above photodisintegration process, an
integral equation of rather similar structure is obtained for the
corresponding amplitude \cite{Barb70,Gibs75},

\begin{equation}
\label{pampl}
 {\cal M} (\vec q \,) = {\cal B} (\vec q\,)
 + \int \! d^{\,3} q' \: {\cal V} (\vec q, \vec q\,')\:{\cal G}_0 (\vec q\,')
\: {\cal M} (\vec q\,') .
\end{equation}

In both equations the kernel is given by an effective neutron-deuteron
potential ${\cal V}$ and an effective free Green function ${\cal G}_0$.
However, in Eq. (2) the inhomogeneity of Eq. (1) is replaced by an off-shell
extension of the photodisintegration amplitude in plane-wave (Born)
approximation,

\begin{equation}
\label{bampl}
 B (\vec q\,) = \: \langle\vec q\,|  \langle \psi_{d}| 
H_{\mbox{\scriptsize em}}|\psi_t 
\rangle \,.
\end{equation}

\noindent
Here, $|{\psi_t} \rangle$ and $| {\psi_d} \rangle$ are the triton and deuteron
states, $|{\vec q\;}\rangle $ is the relative momentum state of the neutron,
$H_{\mbox{\scriptsize em}}$ denotes the electromagnetic operator. In other
words, with this replacement any working program for $n$-$d$ scattering
\cite{Bart87,Janu93} can immediately be applied to the corresponding 
photoprocess.

The results presented in this contribution are obtained by employing,
instead of the {\em W}-matrix representation used in \cite{Schad97a},
 the PEST, BAEST and BBEST potentials as input, however, with an
improved parametrization by Haidenbauer \cite{Haidprivate}.
 The high quality of this input has been demonstrated
in bound-state and scattering calculations
\cite{Haid86b,Corne90a,Park91}.


The electromagnetic operator relevant in the total cross section is, at the low
energies considered, essentially a dipole operator. In the differential cross
section we also have to include  the quadrupole operator.
 According to Siegert's theorem \cite{Sieg37}, they are given by

\begin{equation}
{ {H_{em}^{(1)}}}  \sim  - i \, E_\gamma\,
\sum_{i = 1}^3 e_i  \, r_i \, Y_{1 \lambda}(\vartheta_i, \varphi_i) 
\qquad {\rm and} \qquad
{ {H_{em}^{(2)}}} \sim \frac{E_\gamma^2}{\sqrt{20}}\,
\sum_{i = 1}^3 e_i  \, r_i^2 \, Y_{2 \lambda}(\vartheta_i, \varphi_i) \,.
\end{equation}

\noindent
where, $E_{\gamma}$ denotes  the photon energy,
$r_i$ the nucleon coordinates, $e_i$ the electric charges, and
 $\lambda = \pm 1$ the polarization of the photon. 
The triton wave function in (\ref{bampl}) has been calculated with
 a procedure outlined in Refs. \cite{Schad97a,Cant97a}.

\section{RESULTS}

Table \ref{tabtriton} contains the binding energies and components
of the three-body wave functions obtained with high-rank EST expansions
 compared
to results calculated with the original interactions. The binding
energies for the latter have been calculated with a program by
A. Nogga from Bochum. As shown in Table \ref{tabtriton} the corresponding
results  are in very good agreement, which indicates the
high quality of the EST calculations. Another criterion for the
quality of the wave functions is provided by the norm of the
 difference between the $|\psi_{\rm {EST}} \rangle$, obtained in EST approximation,
and the wave function $|\psi_{2d}\rangle$ obtained in a two-dimensional
calculation for the original potential,
%
$
N = || \Psi_{EST} - \Psi_{2d}|| \,.
$
%
In the cases considered here $N$ is always  $< 10^{-3}$, which in fact
shows the quality of the EST results.
For the further calculations we have projected the wave function on 34 channels
($j \leq 4$), i.e. we have taken into account more than 99.6 \% of the whole
wave function. In our previous calculations we projected on a smaller number of
channels. It turned out that this was not enough to get converged results.

\begin{table}[hbt]
\caption{\label{tabtriton}Binding energies and components of the wave functions.}
\begin{tabular*}{\textwidth}{@{}l@{\extracolsep{\fill}}cccccc}
\hline
\multicolumn{1}{c|}{{{$j \leq $ 2}}}   & {$E_T$ (MeV)} &  P(S)    &  P(S')   &   
  P(P)   &   P(D) \\
\hline 
\multicolumn{1}{l|}{     PEST}      &    -7.369  &  90.138  &  1.420 & 0.0634 &
 8.379  \\
\multicolumn{1}{l|}{    Paris}       &   -7.381   &  90.118  &  1.400 & 0.0644
 &  8.417 \\
\hline
\\
\hline
\multicolumn{1}{l|}{    BAEST}       &   -8.285   &  92.592  &  1.236 & 0.0366
 & 6.135 \\
\multicolumn{1}{l|}{  Bonn {\em A}}  &   -8.294   &  92.595  &  1.232 & 0.0366
 &  6.137  \\
\hline
\\
\hline 
\multicolumn{1}{l|}{     BBEST}      &   -8.088   &  91.614  &  1.189 & 0.0476
 &
7.149  \\
\multicolumn{1}{l|}{   Bonn {\em B}} &   -8.101   &  91.616  &  1.184 & 0.0482
 & 7.152 \\
\hline
\end{tabular*}
\vspace{-2mm}
\end{table}

\begin{figure}[hbt]
\begin{minipage}[t]{75mm}  
\psfig{file=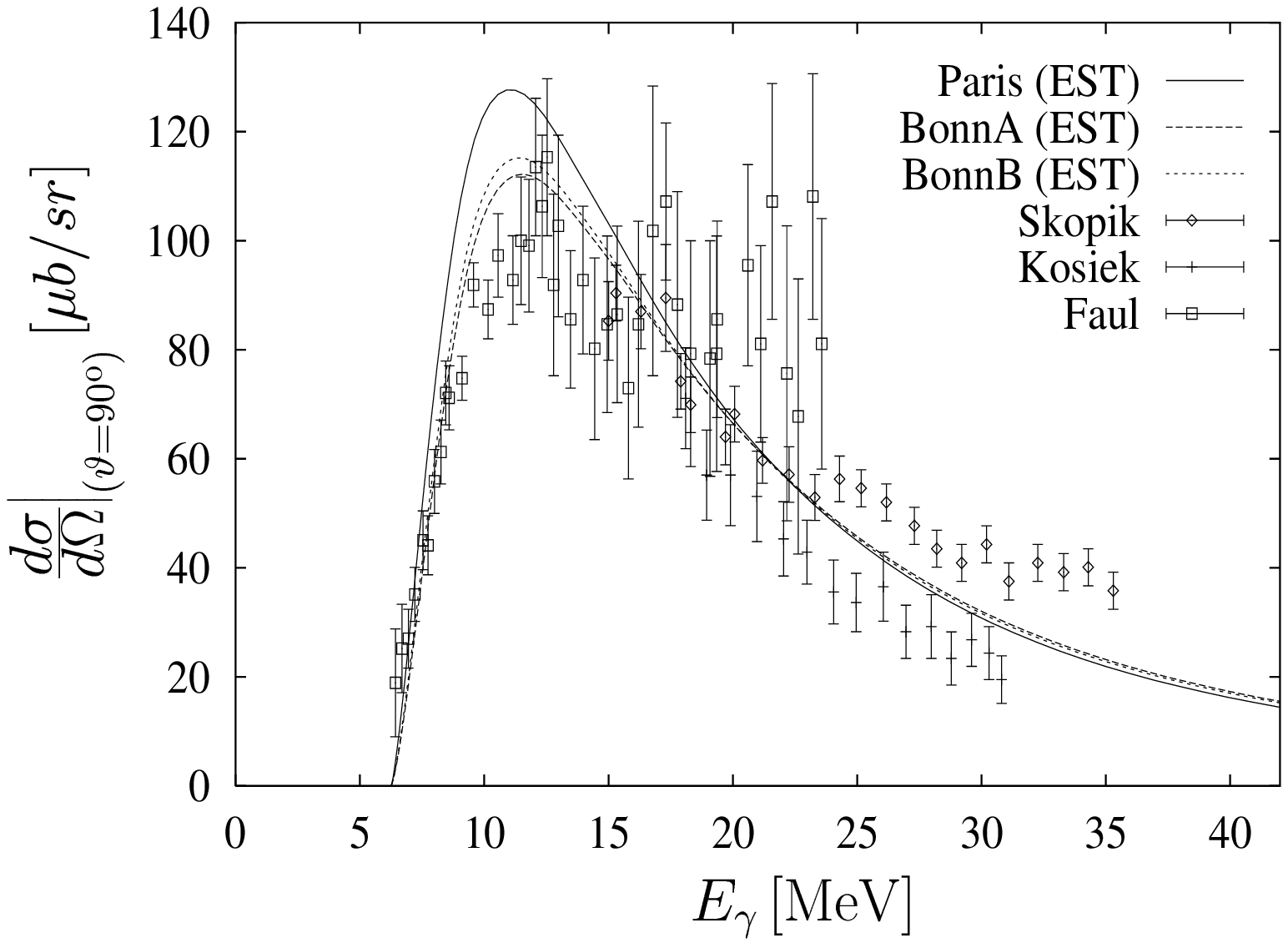,width=49mm}
\vspace{-8mm}
\caption{\label{figtriton} Differential cross section at $\vartheta = 90^{0}$
  for $\gamma + \,^{3}$H$\rightarrow n + d$ obtained with the PEST, BAEST and
  BBEST ($j \leq 2$) potentials. The data are from
  \protect{\cite{Kosi66,Skop81,Faul80}}.}
\end{minipage} 
\hspace{5mm}
\begin{minipage}[t]{75mm}
\psfig{file=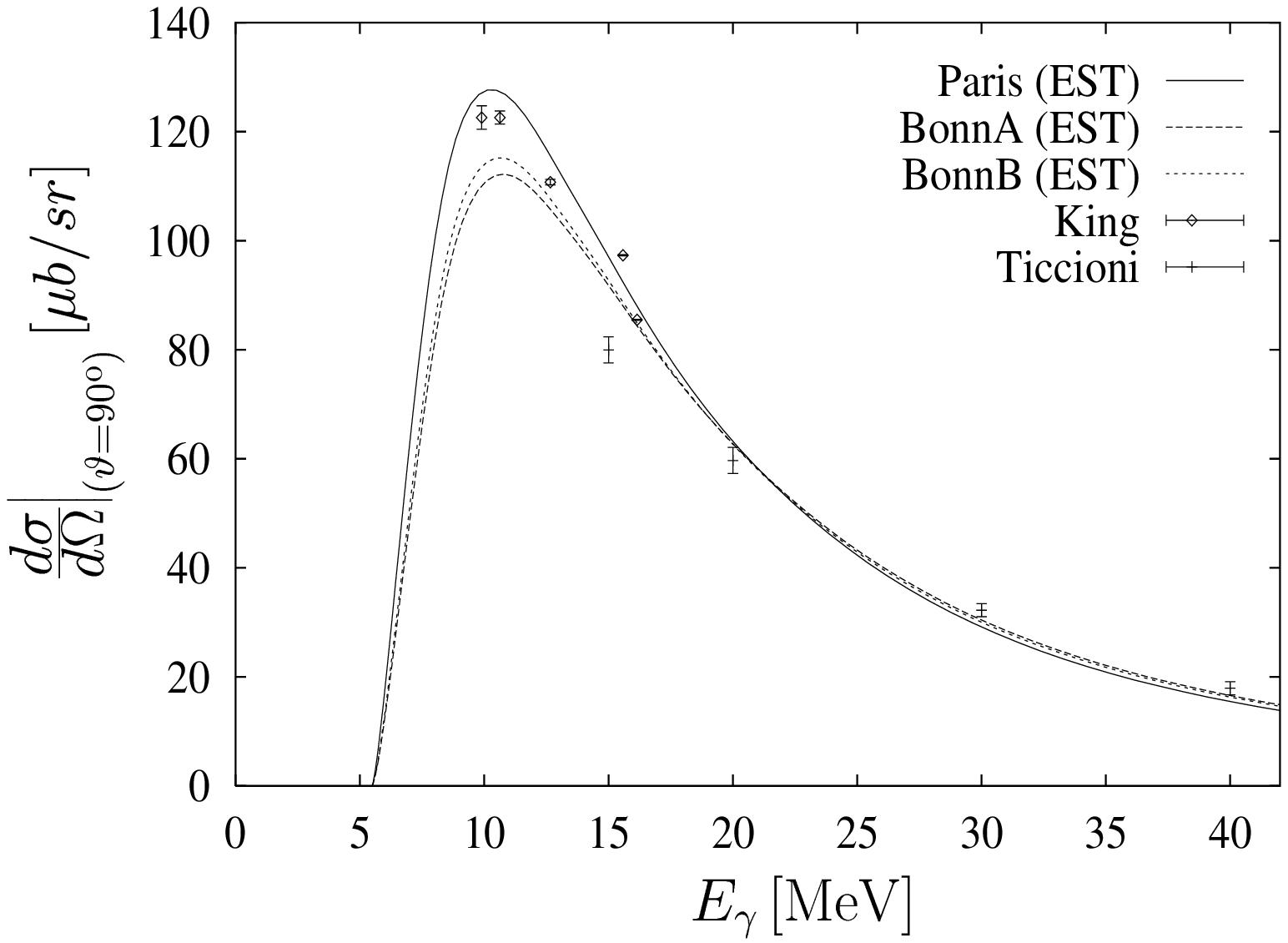,width=49mm}
\vspace{-8mm}
\caption{\label{fighelium} Same as Fig. \ref{figtriton} but for 
the process $\gamma + \,^{3}$He$\rightarrow p + d$. The data are from
\protect{\cite{Ticc73,King84a}}.}
\end{minipage}
\end{figure}

Figure \ref{figtriton} shows the results for the Paris, the Bonn {\em A} and
the Bonn {\em B} potentials compared to the experimental data
\cite{Kosi66,Skop81,Faul80}, the total subsystem angular momentum of the
two-body potential being
restricted to $ j \leq 2$ both in the calculation of $|{\psi_t} \rangle$ and
the treatment of Eq.~(\ref{pampl}). As observed already in our previous
 {\em  W}-matrix approach \cite{Schad97a} there is a considerable
discrepancy between the Paris and Bonn results in the low energy peak region.

\noindent
 Up to 25 MeV the better agreement with the data is achieved for the Bonn
 potentials.  In view of the experimental errors
the relevance of this observation is, of course, somewhat questionable.  At
higher energies the potential dependence vanishes, but in contrast to the
not fully converged results of \cite{Schad97a}, which favored the data
by Skopik et.~al. \cite{Skop81},  our curves lie now almost exactly
between the two sets of data \cite{Kosi66,Skop81}.
More accurate low-energy measurements, therefore, appear highly desirable also
in this energy region.
They would allow us to estimate the quality of the potential models applied in
the current calculations, and to clarify the discrepancies between the data of
Kosiek et~al. \cite{Kosi66}  and Skopik et~al. \cite{Skop81} and our
calculations.

\begin{figure}[hbt]
\centerline{\psfig{file=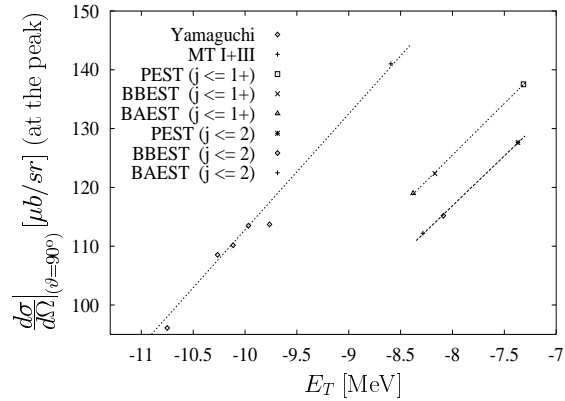,width=49mm}
\phantom{000000000000000000}}
\vspace{-8mm}
\caption{\label{figcorr} Correlation of the peak heights and the binding
  energies for different potentials. The result for Malfliet-Tjon (MT I+III)
 is taken from \cite{Schad97a}.}
\end{figure}

Figure \ref{fighelium} shows the theoretical curves of Fig. \ref{figtriton}
 compared, however,
with the experimental data for $^{3}$He photodisintegration.
 Here, the best agreement is achieved for the  Paris potential.
In both cases the potential which gives a binding energy close
to the experimental values agrees best  with the data for the
 photodisintegration.
To show that there is a correlation between the binding energy and the
height of the peak, we have plotted these values in Figure \ref{figcorr}
for different potentials.
 The results for all potentials with the same number of partial waves  are on
 a straight line.  There is one line for the $s$-wave  MT~I+III potential and
 Yamaguchi potentials with different parameters. The r.h.s. straight line
concerns the Paris, Bonn {\em A} and Bonn {\em B} potentials including
their higher partial wave contributions, in particular the  $p$- and 
$f$-waves. The line in between is the same relationship, however, with
the higher partial waves being switched off.

\vspace{-2mm}

\begin{figure}[hbt]
\begin{minipage}[t]{75mm}  
\psfig{file=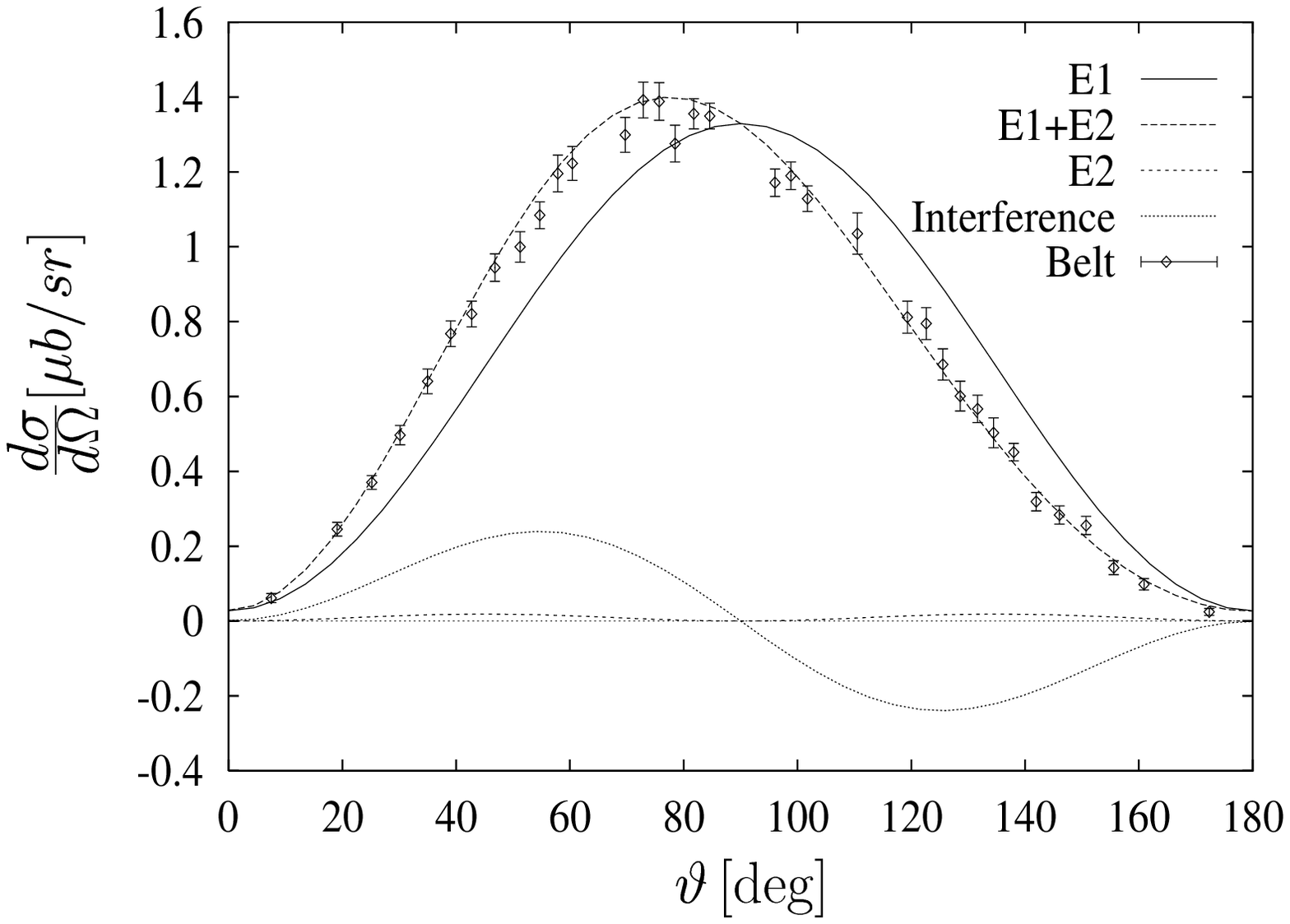,width=49mm}
\vspace{-8mm}
\caption{\label{figpdcap3}$p$-$d$ capture for the PEST potential ($j \leq 2$) 
at  $E$ = 12.1 MeV. $E1$- and $E2$ contributions compared to experimental data 
\protect{\cite{Belt70}}.}
\end{minipage} 
\hspace{5mm}
\begin{minipage}[t]{75mm}
\psfig{file=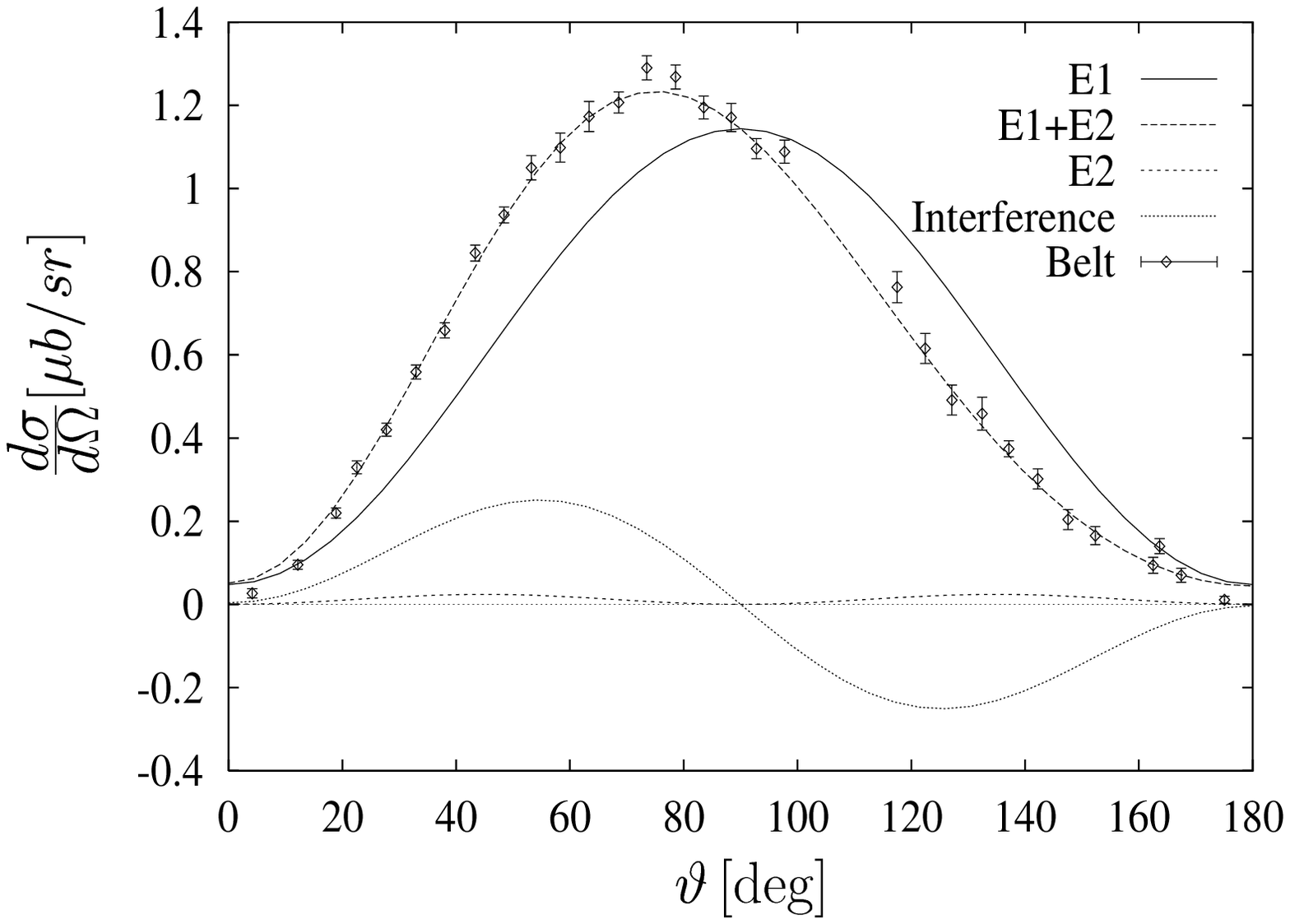,width=49mm}
\vspace{-8mm}
\caption{\label{figpdcap4}$p$-$d$ capture for the PEST potential ($j \leq 2$)
 at $E$ = 15.3 MeV. $E1$- and $E2$ contributions compared to experimental data
 \protect{\cite{Belt70}}.}
\end{minipage}
\vspace{-3mm}
\end{figure}

Figures \ref{figpdcap3} and \ref{figpdcap4} show the differential
cross section for  $p$-$d$ radiative capture. Final-state interaction
and meson exchange currents are incorporated and it is seen that electric
quadrupole $E2$ contributions are needed to achieve the remarkable
agreement between theory and experiment obtained.


\end{document}